\documentclass[10pt,twoside]{article}

\usepackage{a4}

\usepackage{epsfig,subfigure,picinpar}
\usepackage{amsmath,amssymb}
\newlength{\head}

\newlength{\topp}
\newcommand{\bs}[1]{
             \ensuremath{\boldsymbol{#1}}}
\newcommand{\ee}{\ensuremath{\equiv}}
\newcommand{\fr}[1]{
             \ensuremath{\frac{#1}}}
\newcommand{\psibar}{\ensuremath{\overline{\psi}}}
\newcommand{\rhat}{\ensuremath{\boldsymbol{\hat{r}}}}

\newcommand{\p}{\ensuremath{\prime}}

\newcommand{\wk}{\ensuremath{\omega_{\bs{k}}}}

\newcommand{\khat}{\ensuremath{\boldsymbol{\hat{k}}}}

\newcommand{\longr}{\ensuremath{\longrightarrow}}

\begin{document}
\title{Strong and electromagnetic decays 
for excited heavy mesons}
\author{Aksel Hiorth Orsland and H. Hogaasen \\
Department of Physics, University of Oslo,\\
P.O.Box 1048 Blindern, N-0316 Oslo, Norway}
\maketitle
\begin{abstract}
We discuss a model for heavy mesons where the light quark
($u$ or $d$) moves in the colour electric field from a heavy quark
($c$ or $b$) placed in the center of the bag. We calculate energy spectra 
for pionic and photonic transitions from excited states. The transition 
amplitudes and the branching ratios between electromagnetic and pionic 
transitions compares favorable with the limited amount of known experimental
data.
\end{abstract}
\section{Introduction}

Heavy quark spectroscopy is a very interesting and rewarding subject for study.
The discovery of charmonium definitely swept away all doubts the physics 
community had about the existence of quarks as the fundamental building blocks
of hadrons. 

Mesons with one heavy and one light quark is a further excellent laboratory
to test our ideas about strong interactions. These mesons are in a way the
hydrogen atoms of quark physics. As the mass of the heavy quark increases, 
its motion become gradually less and less important and the physical 
properties of the heavy-light, $Q\overline{q}$, meson are more and more 
determined by the dynamics of the light quark.

The discovery of the heavy quark symmetries by Isgur and Wise 
\cite{isgur2,isgur3} and the creation of a heavy quark 
effective theory from QCD\cite{georgi1,eichten1,eichten2}
has been extremely important for the analysis of the physics of heavy hadrons
\cite{neubert1}

Ideally one would like to compute the couplings in the baryon and meson 
Lagrangian from QCD - in time this should be provided by lattice QCD 
calculations. In the meantime model calculations can be useful and one might 
hope that these give us some physical insight for the long distance behavior
of the quark interaction.

\section{The model}

There are many models used in quark physics and we have chosen a variant 
of the M.I.T. bag model that was created by W. Wilcox, O. V. Maxwell and 
K. A. Milton \cite{wilcox} (WWM), at a time when there were little information 
about exited systems made of one heavy and one light quark.

The model is a nice theoretical laboratory, it lends itself to analytical
calculations and it seems to give results that are not too far from 
experimental results. In particular it seems to work well for calculations 
of the Isgur-Wise function \cite{hog2} and to represents an
improvement over results coming from the M.I.T. bag model \cite{sad2}. In the 
WMM-model the heavy quark is placed in the center of the bag and the 
light quark moves in the colour electromagnetic field set up by the heavy 
quark.

The Hamiltonian for the light quark is then
\begin{align}
H&=H_0+H_I\qquad\text{where}\\
H_0&=\bs{\alpha\cdot p}+\beta m +g\bs{t}_{la}V^a,\ \bs{t}_a\ee
\fr{\bs{\lambda_a}}{2}
\ a=1,\ldots ,8 
\quad\text{and}\label{eq:uph}\\
H_{I}&=-g\bs{t}_{la}\bs{\alpha}\cdot\bs{A}^{a}.
\label{eq:ph}
\end{align}
$\bs{t}_a$ are the generators of the $SU(3)_C$ colour group.
The index $l(h)$ refers to the light(heavy) quark.
We have used $A_{a}^{\mu}=(V_{a},\bs{A}_{a})$ and the usual notation
$\alpha_{i}=\gamma^{0}\gamma_{i}$ and $\beta=\gamma^{0}$, where the 
$\gamma$'s are the Dirac matrices and $\bs{\lambda}_a$ the Gellman 
matrices. $V_{a}$ and 
$\bs{A_{a}}$ are the colour electric potentials and vector fields
 respectively produced by the heavy quark.

As gluon selfcouplings are neglected and 
the heavy quark is treated as point like the potential has a Coulomb like 
form :
\begin{equation}
V_{a}=\frac{g\bs{t}_{ha}}{4\pi r}
\end{equation}
Substituting this potential into equation (\ref{eq:uph}) give us :
\begin{equation}
H_{0}=\bs{\alpha}\cdot\bs{p}+\beta m 
+\frac{g^2\bs{t}_l^a\bs{t}_{ha}}{4\pi r}.
\end{equation}
Using the constraint that the meson is a colour singlet, that is 
$\bs{t}_l^a\bs{t}_{ha}=-4/3$ the equation of motion of the light quark in 
the meson rest frame is then simply 
\begin{equation}
H_{0}=\bs{\alpha}\cdot\bs{p}+\beta m -\frac{\xi}{r},\label{eq:h02}
\end{equation}
where $\xi=\frac{4}{3}\alpha_{s}=\frac{4}{3}\fr{g^2}{4\pi}$.

The four component wave function $\psi(r)$ of the light quark (ignoring 
$H_I$) is therefore  the well known solutions for the relativistic Coulomb 
problem. We shall use the notation
\begin{equation}\label{eq:wavefl}
\psi(r)=\begin{pmatrix}g(r)\chi_\kappa^\mu \\
if(r)\chi_{-\kappa}^\mu\end{pmatrix},
\end{equation}
where $\chi_\kappa^\mu$ is the two component spinors describing the 
angular part of the wavefunction.

The energy of the confined light quark is determined by the 
Bogolioubov-MIT boundary condition \cite{bog,chodos:1} which in the 
rest frame of the meson takes the form\linebreak
 $-i(\rhat\cdot\bs{\gamma})\psi=\psi$. 
Substituting equation (\ref{eq:wavefl}) into this equation and using the 
following property $(\bs{\sigma\cdot\rhat})\chi_\kappa^\mu=-\chi_{-\kappa}^\mu$
give us :
\begin{equation}\label{eq:bound}
f(R)+g(R)=0,
\end{equation}
where $R$ is the radius of the spherical bag. The confinement of the light 
quark presumably originating from the gluonic selfcouplings is now taken 
care of by equation (\ref{eq:bound}) and the surface conditions :
\begin{align}
\rhat\cdot\bs{E}^a=0\label{eq:bounde1} \\
\rhat\times\bs{B}^a=0\label{eq:boundb1}
\end{align}

\begin{figure}[tb]
\begin{center}
   \epsfig{file=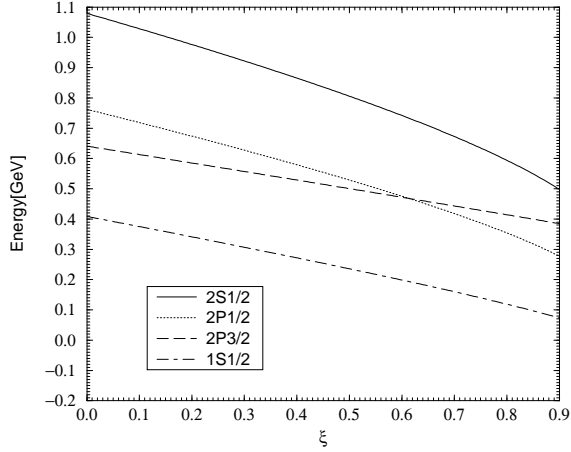,height=7cm}
\caption{Energy levels inside a bag with radius $R=5GeV^{-1}$.}
\label{fig:enlev4}
\end{center}
\end{figure}

The vector fields $\bs{A}_a$ that are set up by the heavy quark 
and fulfill the boundary conditions (\ref{eq:boundb1}) are 
\begin{equation}
\bs{A}_a=\fr{1}{4\pi}(\fr{\bs{m}_a\times\rhat}{r^3}+
\fr{\bs{m}_a\times\rhat}{2R^3}),
\end{equation}
where $\bs{m}_a$ is the colour magnetic moment(s) of 
the heavy quark :
\begin{equation}
\bs{m}_a=\fr{g\bs{t}_{ha}}{M_h}\bs{S}.
\end{equation}
$M_h$ is the mass of the heavy quark and $\bs{S}$ its spin operator.
$H_I$ now takes the form :
\begin{equation}
H_I=\fr{\alpha_s\bs{t}_{la}{\bs{t}_h}^a}{M_h}\bs{S}\cdot 
(\bs{\alpha}\times\rhat)
(\fr{1}{r^2}+\fr{r}{2R^3})
\end{equation}
The contribution of $H_I$ to the energy is calculated perturbatively and 
the hyperfine splitting energy to first order in $\alpha_s$ is
\begin{equation}
E^1_I=\fr{8}{3}\fr{\alpha_s}{M}\fr{\kappa}{4\kappa^2-1}
(F(F+1)-J(J+1)-\fr{3}{4})
\frac{1}{N}\int_0^R\,dr(2+\fr{r^3}{R^3})f(r)^*g(r).\label{eq:i}
\end{equation}
$N$ is the normalization of the wavefunction, 
$N\ee\int_0^R\,drr^2(|f(r)|^2+|g(r)|^2)$. 
Here $F$ is the total angular momentum of the mesonic system and $J$ is the 
light quark (total) angular momentum. From equation (\ref{eq:i}) we 
see for $M_h\rightarrow\infty$ then  
$E_I^1\rightarrow 0$; this is the heavy quark limit.

The mass functional for a heavy meson described 
in our bag model will be :
\begin{equation}
M=M(R)=E_{Vol}+E_{Zero}+m_Q+E_q\label{eqm(r)}
\end{equation}
where $E_{Vol}=\fr{4\pi}{3}BR^3$ is the energy needed to create 
a bag in vacuum , $E_{Zero}$ is the zero point energy proportional to 
$1/R$, $m_Q$ is the heavy quark mass and $E_q$ is the light quark energy
$E_q\ee\sqrt{p_q^2+m_q^2}+E_I^1$, where $E_I^1$is the hyperfine 
splitting energy to first order given in equation (\ref{eq:i}).
We will say more about the first two terms later. 
For now we only note that if the radii of two  mesons 
with same flavour of the heavy 
quark is kept constant, then the mass difference between them are given
by the following formula :
\begin{equation}
\Delta M=E_q(nL_J)-E_q(n^\prime L^\prime_{J^\prime}).
\end{equation}
This means that if the radii of two mesons do not differ too much, then
the difference between the energy levels is directly 
related to the mass difference of the two mesons.

It is of interest first to see how the energy
levels of the meson are ordered in the 
heavy quark limit when the colour electric central potential increases, these 
are shown in fig. \ref{fig:enlev4}.

As we can see an increase in the central field from the special case where the 
light quark moves freely in the bag, reduces the mass of the $P$ states 
- and even make them cross. 
The odd parity states stay however roughly half way
between the ground state and the first excited $S$-state.
This will lead to some difficulties when we try to fit the spectrum of heavy 
meson states.

The breaking of the heavy quark limit is given by the spin-spin interaction and
in our model by the term $E_I^1$, given in equation (\ref{eq:i}).
It should be noted that this term is dependent of both 
$1/M$ and $\alpha_s$ which 
determines the strength of the central four vector potential the light quark 
moves in. In this respect our model is more constraining than most models
where the interquark central potential is unrelated to the 
strength of the spin-spin
and spin-orbit interaction.
\begin{figure}[htb]
\begin{center}
   \epsfig{file=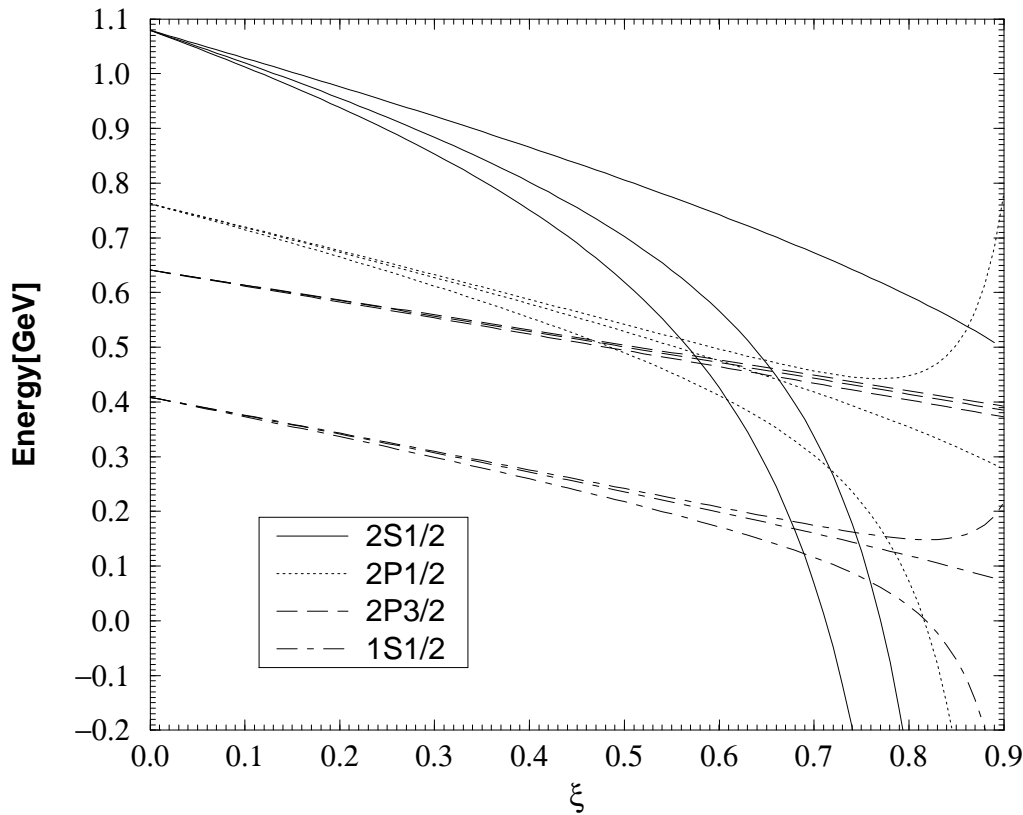,width=\linewidth}
   \caption{The energy of a (massless) light quark inside a Coulomb bag with
radius $R=5GeV^{-1}$ and a mass of the heavy quark $m_b=4730 MeV$}
\label{fig:enhq}
\end{center}
\end{figure}

In figure \ref{fig:enhq} we have plotted the energy levels for the light quark 
in the case where $M=m_b$ for constant $R$.
In this graph we have also plotted the heavy quark limit, the energy level
 above the heavy quark limit is for each pair of states 
where the spin of the light quark and the heavy quark couples to $S=1$ 
and the level below for the case where the spins couples to $S=0$.

We see that for a given heavy quark multiplet, the induced splitting of the 
formerly degenerate states with the same angular momentum $J$ of the light 
quark, but with different angular momentum $F$ for the meson, is a highly 
nonlinear function of $\alpha_s$. Only for $\alpha_s$ smaller than $0.2$ 
can the hyperfine splitting with a reasonable approximation be taken as
a linear function of $\alpha_s$ as it is in the nonrelativistic quark 
model treatment.

From figure \ref{fig:enhq} we also note that the hyperfine splitting 
(for finite $\xi$) increases as we go up in light quark excitations. This 
is quite opposite from the situation in the hydrogen atom and is a 
reflection of the bag models abrupt confinement.

For the charm sector only for the two lowest states it is reasonable 
to calculate the hyperfine splitting perturbatively, this is because
of the much smaller mass of the $c$-quark.

\section{The mass functional for heavy mesons}

\begin{figure}
\begin{center}
   \epsfig{file=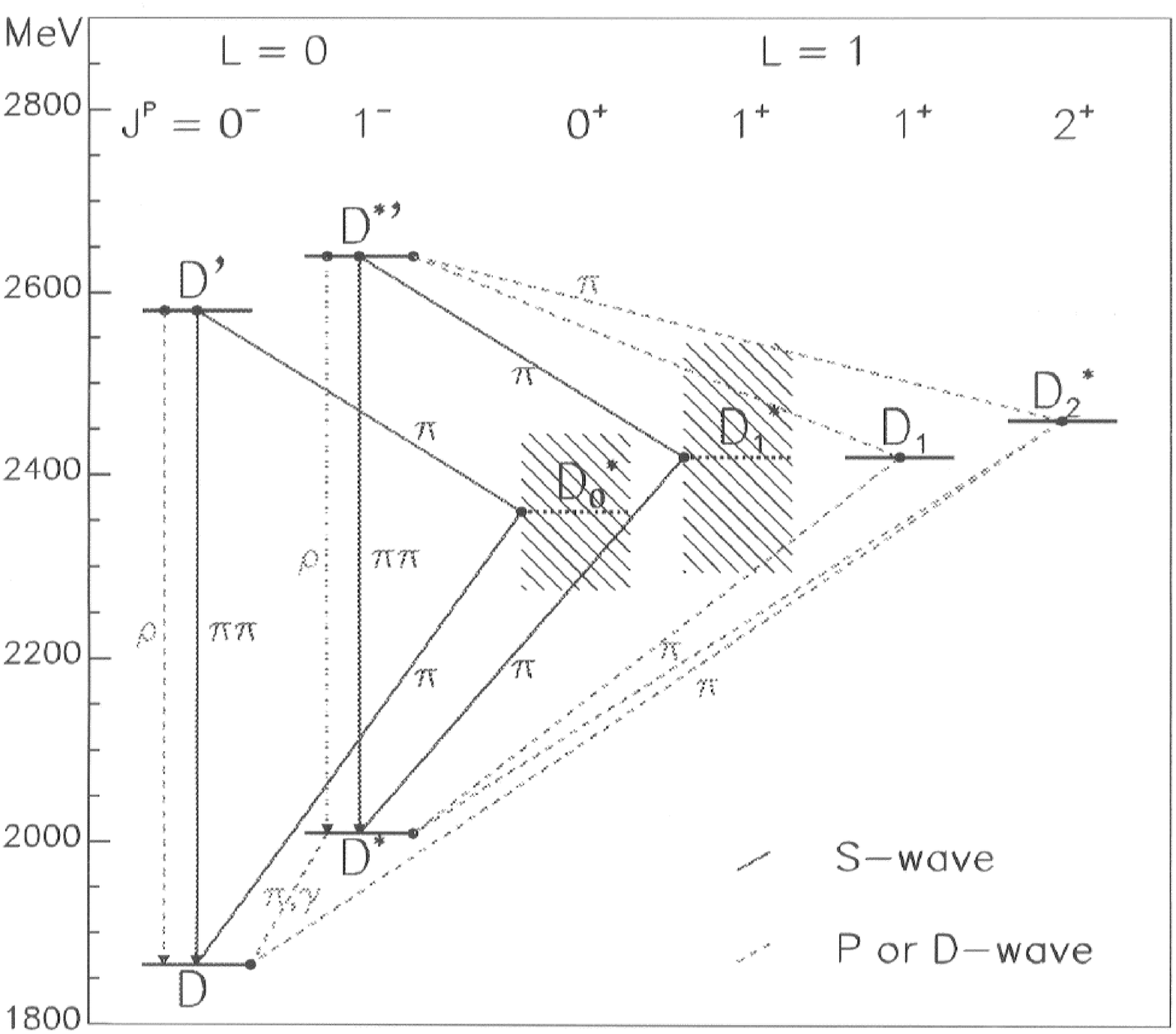,width=\linewidth}
   \caption{$D$ meson spectra and transition lines.}\label{fig:dmeson}
\end{center}
\end{figure}

\begin{figure}
\begin{center}
   \epsfig{file=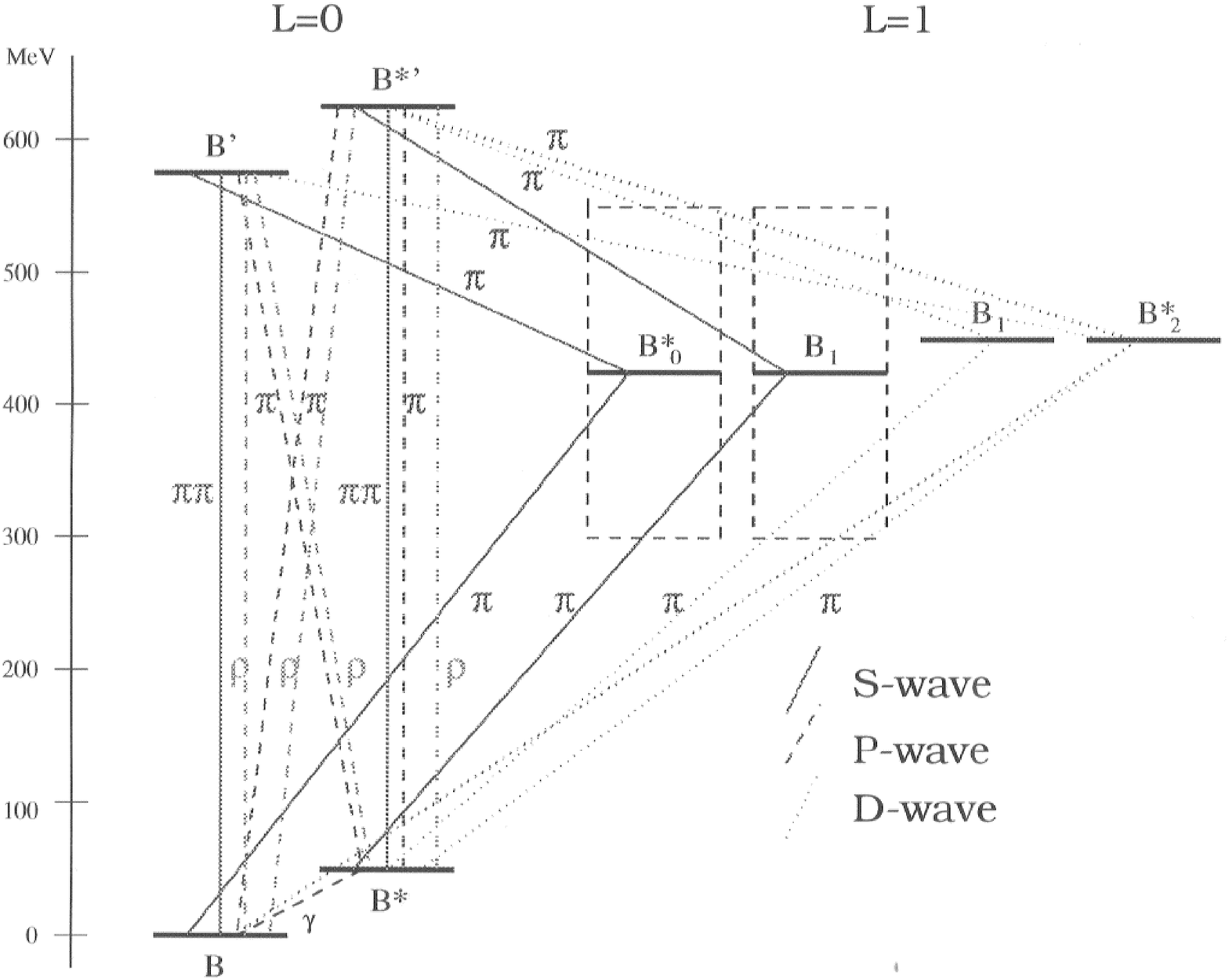,width=\linewidth}
   \caption{$B$ meson spectra and transition lines.}\label{fig:bmeson}
\end{center}
\end{figure}

In the previous section we looked at the qualitatively features of the 
spectra of the heavy mesons. Now we
will try to reproduce the quantitatively measured masses. In 
figure \ref{fig:dmeson} and \ref{fig:bmeson} we have showed 
the observed spectra of the $D$ and $B$ mesons. The mass 
formula for heavy mesons in our model is :
\begin{align}
M&=\fr{4}{3}BR^3+\fr{C}{R}+m_Q+\sqrt{{p_q}^2+m_q}
+\frac{8}{3}\frac{\alpha_s}{M}
\frac{\kappa}{4\kappa^2-1}(F(F+1)-j(j+1)-\frac{3}{4})\cal{I}\label{eq:massf} \\
\intertext{where $\cal{I}$ is an integral over the radial wavefunctions}
\cal{I}&=\frac{1}{N}\int_0^R\,dr(2+\fr{r^3}{R^3})f(r)^*g(r)\label{eq:mint}
\end{align}

When calculating the Coulomb potential in classical electrodynamics it is 
common to choose 
\begin{equation}
V(r)=0\qquad\text{for}\qquad r\longr\infty
\end{equation}
Inspired by this we adjust the potential inside the bag 
to be zero at the surface of the bag \cite{hog}:
\begin{equation}
V(r)=-\fr{\xi}{r}\longr -\xi(\fr{1}{r}-\fr{1}{R})
\end{equation}
Now the value of the potential at the bag surface will be zero, 
independent of the radii of the different mesons. Because of this 
transformation the light quark gets a contribution $\xi/R$ to the energy.
This we include in the light quark energy.

The second term, $C/R$, in equation (\ref{eq:massf}) is supposed 
to represent the 
zero point energy in the bag. When one quantizes a radiation field there 
will always be an infinite zero point energy term, but since physical 
quantities often are energy 
differences, the zero point energy falls out. However, when 
the quantization is carried out in a finite cavity, as in our model, there 
will be additional pieces of the zero point energy 
which depends on the size of the 
cavity. This is represented by the term $C/R$. The constant $C$ can be 
calculated if we believe that the zero point energy give rise to the Casimir
energy. The Casimir energy inside a perfectly uncharged spherical shell 
has been calculated by K.A.Milton, L.L.DeRaad, Jr. and J.Schwinger in 
\cite{milton}. They obtained a value $E=0.09235/(2R)$. To find the value 
in our model we simply have to multiply the value by eight, because
 there are eight gluonic radiation fields :
\begin{equation}
C\simeq 0.37
\end{equation}

We have determined the masses by minimizing the mass functional by the 
relation :
\begin{equation}
\fr{\partial M}{\partial R}=0\label{eq:minm}
\end{equation}
It turns out that the integral in equation (\ref{eq:mint}) goes as 
\begin{equation}
\fr{1}{m_QR^2}
\end{equation}
Since $p_q\sim 1/R$ it is clear that when $R\longr 0$, 
$\cal{I}$ will dominate. If the 
factor in front of the hyperfine splitting is negative, equation 
(\ref{eq:massf})
will for some choice of the parameters have no finite minima.
When the radius become small we can not neglect the 
repulsion of the heavy quark and the model becomes meaningless. This 
is a well known problem and the usual way of dealing with this is 
 to argue that the divergence 
will disappear when we calculate higher orders correction. 
The procedure then is to minimize the energy with respect to 
$R$ \emph{before} adding the hyperfine term \cite{detar}. In our model
states with same $J$ and $L$ will then have equal radii.

\section{Heavy meson masses}

First of all we have to determine the parameters in the model. There 
are four parameters the strong coupling constant($\alpha_s$), the bag 
constant($B$), the heavy meson mass($m_Q$) and the light quark mass($m_q$).
There are of course many ways of determining the parameters, we have chosen 
too look at the $B$ mesons. We have chosen to determine 
$\alpha_s, B, m_b$ from the 
observed masses of the $B, B^*$ and $B^\p$ mesons, and since the $u$ and $d$
quark have a small mass we have assigned to them a (rather unimportant) 
mass of $10 MeV$.
The calculated results 
are shown in table \ref{tab:masstab1}

\begin{table}[htb]
\begin{center}
\begin{tabular}{|l|c|c|c|c|c|c|}                                 \hline\hline
\multicolumn{2}{|c|}{Experimental}&\multicolumn{5}{c|}{Theoretical} \\
\multicolumn{2}{|c|}{Results}&\multicolumn{5}{c|}{Results} \\ \hline
         & Mass[$MeV$] & Mass[$MeV$]  & $F^P$ & State & Radius[$GeV^{-1}]$ & $\mu_m/\mu_N$ \\ \hline\hline
$B$      & $5279.4\pm 2.2\ \cite{rev}$   & $5279$  & $0^-$ & $1S_{1/2}$ & $3.94$ &$0$\\ \hline  
$B^*$    & $5324.8\pm 1.8\ \cite{rev}$   & $5325$  & $1^-$ & $1S_{1/2}$ & $3.94$ &$1.52e_q+0.203e_Q$\\ \hline  
$B_0$    & $?$   & $5592$  & $0^+$ & $2P_{1/2}$ & $4.50$ &$0$ \\ \hline               
$B_1$    & $?$   & $5671$  & $1^+$ & $2P_{1/2}$ & $4.50$ &$0.625e_q+0.203e_Q$\\ \hline       
$B_1$    & $5725\ \cite{opal}$   & $5623$  & $1^+$ & $2P_{3/2}$ & $4.47$ &$1.93e_q+0.101e_Q$\\ \hline       
$B_2$    & $5737\ \cite{opal}$   & $5637$  & $2^+$ & $2P_{3/2}$ & $4.47$ &$2.33e_q+0.203e_Q$\\ \hline   
$B^{\prime}$ & $5859\ \cite{weiser}$   & $5859$  & $0^-$ & $2S_{1/2}$ & $4.90$&$0$ \\ \hline
$B^{\p *}$ & $?$   & $5967$  & $1^-$ & $2S_{1/2}$ & $4.90$ &$0.597e_q+0.203e_Q$\\ \hline\hline
\end{tabular}
\caption{The parameters are $B^{1/4}=161MeV, \xi =0.538\  (\alpha_s=0.404), m_u=m_d=10MeV$ 
and $m_b=4627MeV$.$e_q=e_Q=-1/3$ for $d$ and $b$ quarks and 
$e_q=e_Q=2/3$ for $u$ and $c$ quarks.}\label{tab:masstab1}
\end{center}
\end{table}

In table \ref{tab:masstab1} there are only given uncertainties for the 
$1S_{1/2}$ states. This is because the value of the masses for the $2P_{3/2}$
and the lowest $2S_{1/2}$ states was found in \cite{opal} and \cite{weiser}
as a fit to the experimental data and it is not quite clear how to put 
errors to these numbers.

Unfortunately we see that the calculated masses for the $B_1$ and $B_2$ mesons 
are almost exactly $100MeV$ below the experimental values. However, we 
see that the splitting between the two $P_{3/2}$ mesons is quite well 
described. One may wonder 
if there are another set of parameters which will give us a good fit to 
all the experimental masses listed in table \ref{tab:masstab1}. 
We believe that this
is not so. The parameter in the system which has most influence 
on the spectra is $\xi$. In figure \ref{fig:enlev4} we plotted the 
energy levels for $\xi$ between 0 and 0.9. The main problem is that 
the $P$ states does not move much in direction of the 
$2S_{1/2}$ state when $\xi$ increase. The boundary condition (\ref{eq:bound})
clearly gives a wrong level splitting. A hope for the model would be 
that the state we used as $B^\p$ is not the right one. If we used the $B_1$
and $B_2$ states (together with $B$ and $B^*$) as a normalization 
the resulting $B^\p$ would have a mass around $6020 MeV$.

Now we have the value of the strong coupling, the bag constant and the 
$b$ quark mass. In order to calculate the $D$ and $B_s$ meson masses we 
have to find the $c$ and $s$ quark masses. This has been done by demanding 
that :
\begin{align}
m(D_{Exp.})-m(D_{Teor.})&\approx m(D^*_{Theor})-m(D^*_{Exp.}) \\
\intertext{and the s-quark mass by demanding that:} 
m({B_s}_{Exp.})-m({B_s}_{Theor.})&\approx m({B^*_s}_{Theor.})-m({B^*_s}_{Exp.})
\end{align}
The results are listed in table \ref{tab:masstab2}.
\begin{table}[htb]
\begin{center}
\begin{tabular}{|l|c|c|c|c|c|c|}                                 \hline\hline
\multicolumn{2}{|c|}{Experimental}&\multicolumn{5}{c|}{Theoretical} \\
\multicolumn{2}{|c|}{Results}&\multicolumn{5}{c|}{Results} \\\hline
         & Mass[$MeV$] & Mass[$MeV$]  & $F^P$ & State & Radius[$GeV^{-1}]$ &$\mu_m/\mu_N$\\ \hline\hline
$D$      & $1866.9\pm 0.7\ \cite{rev}$   & $1854$  & $0^-$ & $1S_{1/2}$ & $3.94$ &$0$\\ \hline
$D^*$    & $2008.4\pm 0.7\ \cite{rev}$   & $2022$  & $1^-$ & $1S_{1/2}$ & $3.94$ &$1.52e_q+0.725e_Q$\\ \hline  
$B_s$    & $5369.6\pm 2.4\ \cite{rev}$   & $5363$  & $0^-$ & $1S_{1/2}$ & $3.91$ &$0$\\ \hline  
$B_s^*$  & $5416.3\pm 3.3\ \cite{rev}$   & $5424$  & $1^-$ & $1S_{1/2}$ & $3.91$ &$1.52e_q+0.203e_Q$\\ \hline 
$B_{0_{s}}$ & $?$ & $5667$  & $0^+$ & $2P_{1/2}$ & $4.47$                        &$0$\\ \hline 
$B_{1_{s}}$ & $?$ & $5737$  & $1^+$ & $2P_{1/2}$ & $4.47$                        &$0.619e_q+0.203e_Q$\\ \hline       
$B_{1_{s}}$ & $5874\ \cite{opal}$ & $5718$  & $1^+$ & $2P_{3/2}$ & $4.45$        &$1.93e_q+0.101e_Q$\\ \hline 
$B_{2_{s}}$ & $5886\ \cite{opal}$ & $5732$  & $2^+$ & $2P_{3/2}$ & $4.45$        &$2.31e_q+0.203e_Q$\\ \hline   
$B^{\prime}_s$ &  ?   & $5887$  & $0^-$ & $2S_{1/2}$ & $4.87$                    &$0$ \\ \hline
$B^{\p *}_s$ &  ?   & $6008$  & $1^-$ & $2S_{1/2}$ & $4.87$                      &$0.593e_q+0.203e_Q$\\ \hline\hline
\end{tabular}
\caption{The parameters are $B^{1/4}=161MeV, \xi =0.538\ (\alpha_s=0.404)
, m_b=4627MeV$, $m_c=1294MeV$ and $m_s=231MeV$. $e_q=e_Q=-1/3$ for $d$ and $b$ quarks and 
$e_q=e_Q=2/3$ for $u$ and $c$ quarks.}\label{tab:masstab2}
\end{center}
\end{table}

\section{Decays}
\begin{table}[tb]
\begin{center}
\begin{tabular}{|c|l|c|c|}\hline\hline
$J_\alpha^{P\alpha}\longr J_\beta^{P\beta}\pi$&$\alpha\longr\beta\pi$ &Teor.[$MeV$] &$g_{\alpha\beta\pi}$  \\ \hline\hline
$0^-\longrightarrow 1^-\pi$ & $B^{\p}\longrightarrow B^*\pi^+$          & $84.3$ &$42.8$ \\ \cline{2-4}
                             & $B^{\p}\longrightarrow B^*\pi^0$         & $42.2$ &$30.1$\\ \hline
$0^-\longrightarrow 0^+\pi$ & $B^{\p}\longrightarrow B_0\pi^+$          & $45.3$ &$23.0$\\ \cline{2-4}
                             & $B^{\p}\longrightarrow B_0\pi^0$         & $22.7$ &$16.2$\\ \hline
$2^+\longrightarrow 1^-\pi$ & $B_2\longrightarrow B^*\pi^+$             & $9.17$ &$55.8$\\ \cline{2-4}
                             & $B_2\longrightarrow B^*\pi^0$            & $4.67$ &$39.4$\\ \hline
$2^+\longrightarrow 0^-\pi$ & $B_2^+\longrightarrow B^0\pi^+$           & $9.74$ &$43.5$\\ \cline{2-4}
                             & $B_2^0\longrightarrow B^0\pi^0$          & $4.94$ &$30.7$\\ \cline{2-4}
                             & $B_2^+\longrightarrow B^+\pi^0$          & $4.98$ &$30.7$\\ \cline{2-4}
                             & $B_2^0\longrightarrow B^+\pi^-$          & $9.82$ &$43.5$\\ \hline
$1^+(P_{3/2})\longrightarrow 1^-\pi$ & $B_1\longrightarrow B^*\pi^+$    & $13.3$ &$72.8$\\ \cline{2-4}
                             & $B_1\longrightarrow B^*\pi^0$            & $6.71$ &$51.1$\\ \hline
$1^+(P_{1/2})\longrightarrow 1^-\pi$ & $B_1\longrightarrow B^*\pi^+$    & $92.7$ &$27.1$\\ \cline{2-4}
                             & $B_1\longrightarrow B^*\pi^0$            & $46.1$ &$19.0$\\ \hline
$0^+\longrightarrow 0^-\pi$ & $B_0^+\longrightarrow B^0\pi^+$           & $93.7$ &$28.5$\\ \cline{2-4}
                             & $B_0^0\longrightarrow B^0\pi^0$          & $46.8$ &$20.1$\\ \cline{2-4}
                             & $B_0^+\longrightarrow B^+\pi^0$          & $46.9$ &$20.1$\\ \cline{2-4}
                             & $B_0^0\longrightarrow B^+\pi^-$          & $93.7$ &$28.5$\\ \hline
$1^-\longrightarrow 0^-\pi$ & $D^{*+}\longrightarrow D^0\pi^+$          & $6.01\ 10^{-2}$ &$17.1$\\ \cline{2-4}
                             & $D^{*+}\longrightarrow D^+\pi^0$         & $2.72\ 10^{-2}$ &$12.1$\\ \cline{2-4}
                             & $D^{*0}\longrightarrow D^0\pi^0$         & $3.88\ 10^{-2}$ &$12.1$\\ \hline\hline
\end{tabular}
\end{center}
\caption{Pion decays}\label{tab:ratetab2}
\end{table}

We have looked at electromagnetic and pionic transitions between mesons 
listed in table \ref{tab:masstab1} and \ref{tab:masstab2}. The pionic 
transitions are calculated using the surface coupling version of the chiral
bag model\cite{thomas}. In this model a pion field carries the axial 
current outside the bag produced by the light ($u$ or $d$) quark inside 
the bag. This makes the model chirally symmetric for massless $u$ and 
$d$ quarks and the interaction between the bag and the pion field 
is given by :
\begin{equation}
{\cal L}_{\text{int}}=-\fr{i}{2f_\pi}\psibar\gamma_5\bs{\tau\cdot\phi}\psi
\Delta_s,
\end{equation}
$\tau_i$ are the Pauli isospin matrices and $\bs{\phi}$ an isovector 
representing the pion field. $\Delta_s$ is a covariant surface delta-function.

The calculation of transitions involving pions is then straight forward, 
some expressions for pionic transitions are listed 
below :
\begin{align}
0^-\longrightarrow 1^-0^-&\begin{cases}&V_{fi}=\fr{-i}{f_\pi}
\sqrt{\fr{\pi}{3V\wk}}P(R)
\left( Y_1^{-1}(\khat )+Y_1^1(\khat )+Y_1^0(\khat )\right)C_\pi \\ 
&\Gamma (B^\p\longr B^*\pi )=\fr{1}{4\pi}\fr{k}{f_\pi^2}{|P(R)C_\pi|}^2
\end{cases}\label{eq:2s}\\
0^-\longrightarrow 0^+0^-&\begin{cases}&V_{fi}=\fr{-1}{f_\pi}
\sqrt{\fr{\pi}{V\wk}}S(R)Y_0^0(\khat )C_\pi\\ 
&\Gamma (B^\p\longr B_0\pi )=\fr{1}{4\pi}\fr{k}{f_\pi^2}{|S(R)C_\pi|}^2
\end{cases}\\
2^+\longrightarrow 0^-0^-&\begin{cases}&V_{fi}=-\fr{1}{f_\pi}
\sqrt{\fr{2\pi}{5V\wk}}D(R)Y_2^2(\khat )C_\pi \\ 
&\Gamma (B_2\longr B\pi ) =\fr{1}{10\pi}\fr{k}{f_\pi^2}{|D(R)C_\pi|}^2
\end{cases}\\
2^+\longrightarrow 1^-0^-&\begin{cases}&V_{fi}=\fr{1}{f_\pi}
\sqrt{\fr{2\pi}{5V\wk}}D(R)
\left( Y_2^2(\khat )-\fr{1}{\sqrt{2}}Y_2^1(\khat )\right)C_\pi \\ 
&\Gamma (B_2\longr B^*\pi )=\fr{3}{20\pi}\fr{k}{f_\pi^2}{|D(R)C_\pi|}^2
\end{cases}\label{eq:2p3/2}\\
1^+\longrightarrow 1^-0^-&\begin{cases}&V_{fi}=\fr{1}{f_\pi}
\sqrt{\fr{\pi}{10V\wk}}D(R)
\left( \sqrt{6}Y_2^2(\khat )-\sqrt{3}Y_2^1(\khat )+Y_2^0(\khat )\right)C_\pi\\ 
&\Gamma(B_1(P_{3/2})\longr B^*\pi )=\fr{1}{4\pi}\fr{k}{f_\pi^2}{|D(R)C_\pi|}^2
\end{cases}\label{eq:p32} \\
1^+\longrightarrow 1^-0^-&\begin{cases}&V_{fi}=\fr{1}{f_\pi}
\sqrt{\fr{\pi}{V\wk}}S(R)Y_0^0(\khat )C_\pi \\ 
&\Gamma (B_1(P_{1/2})\longr B^*\pi )=\fr{1}{4\pi}\fr{k}{f_\pi^2}{|S(R)C_\pi|}^2
\end{cases}\label{eq:p12} \\
0^+\longrightarrow 0^-0^-&\begin{cases}&V_{fi}=-\fr{1}{f_\pi}
\sqrt{\fr{\pi}{V\wk}}S(R)Y_0^0(\khat )C_\pi \\ 
&\Gamma (B_0\longr B\pi )=\fr{1}{4\pi}\fr{k}{f_\pi^2}{|S(R)C_\pi|}^2
\end{cases}\label{eq:p120} 
\end{align}
\begin{align}
1^-\longrightarrow 0^-0^-&\begin{cases}&V_{fi}=\fr{-i}{f_\pi}
\sqrt{\fr{\pi}{3V\wk}}P(R)Y_1^1(\khat )C_\pi\\ 
&\Gamma (D^*\longr D\pi )=\fr{1}{12\pi}\fr{k}{f_\pi^2}{|P(R)C_\pi|}^2.
\end{cases}\label{eq:1s} \\
\intertext{In these formulae :}
S(R)\ee&R^2\left( f_\beta^*(R)g_\alpha (R)+
g_\beta^*(R)f_\alpha (R)\right)j_0(kR)\qquad\text{(S-wave.)} \\
P(R)\ee& R^2\left( f_\beta^*(R)g_\alpha (R)+
g_\beta^*(R)f_\alpha (R)\right)j_1(kR)\qquad\text{(P-wave.)} \\
D(R)\ee& R^2\left( f_\beta^*(R)g_\alpha (R)+
g_\beta^*(R)f_\alpha (R)\right)j_2(kR)\qquad\text{(D-wave.)} \\
\intertext{The index $\alpha(\beta)$ refers to the initial(final) meson.}
C_\pi&\ee\begin{cases} 1\quad &\text{for}\quad \pi^\pm \\
                      \fr{1}{\sqrt2}\quad &\text{for}\quad \pi^0.
\end{cases}
\end{align}
The above transition rates 
(\ref{eq:2s})-(\ref{eq:1s}) have been numerically 
calculated and are listed in table \ref{tab:ratetab2}.

In addition to the partial widths listed in table \ref{tab:ratetab2} we 
have also calculated the coupling constants (to the right in table 
\ref{tab:ratetab2}), by using the following definition :
\begin{equation}\label{eq:defcoupl}
g_{\alpha\beta\pi}\ee\sqrt{\fr{\Gamma(\alpha\longr\beta\pi)
 24\pi M_\alpha^2}{k^{2L+1}}}
\end{equation}
$k$ is the pion momenta and $M_\alpha$ the mass of the decaying particle.
The dimension of the couplings goes as 
$[GeV]^{-L+1}$, where $L$ is the relative angular momentum between the 
decay products. Only $L=1$ transitions such as $1^-\longr 0^-\pi$ 
are then dimension less by the definition (\ref{eq:defcoupl}).
\begin{table}[htb]
\begin{center}
\begin{tabular}{|c|l|c|}\hline\hline
$J_\alpha^P\longr J_\beta^P\pi$&$\alpha\longr\beta\pi$ &Teor.[$keV$]   \\ \hline\hline
$0^-\longrightarrow 2^++1^-$ & $B^{**0}\longrightarrow B_2^0\gamma$     & $2.62\ 10^{-3}$ \\ \cline{2-3}
                             & $B^{**+}\longrightarrow B_2^+\gamma$     & $1.05\ 10^{-2}$ \\ \hline
$0^-\longrightarrow 1^++1^-$ & $B^{**0}\longrightarrow B_1^0\gamma$     & $1.77\ 10^{1}$ \\ \cline{2-3}
                             & $B^{**+}\longrightarrow B_1^+\gamma$     & $7.07\ 10^{1}$ \\ \hline
$0^-\longrightarrow 1^-+1^-$ & $B^{**0}\longrightarrow B^{*0}\gamma$    & $1.20\ 10^{-1}$ \\ \cline{2-3}
                             & $B^{**+}\longrightarrow B^{*+}\gamma$    & $7.89\ 10^{-1}$ \\ \hline
$0^-\longrightarrow 0^-+1^-$ & $B^{**0}\longrightarrow B^0\gamma$       & $0$ \\ \cline{2-3}
                             & $B^{**+}\longrightarrow B^+\gamma$       & $0$ \\ \hline
$2^+\longrightarrow 1^++1^-$ & $B_2^0\longrightarrow B_1^0\gamma$       & $6.26\ 10^{-2}$ \\ \cline{2-3}
                             & $B_2^+\longrightarrow B_1^+\gamma$       & $2.50\ 10^{-3}$ \\ \hline
$2^+\longrightarrow 1^-+1^-$ & $B_2^0\longrightarrow B^{*0}\gamma$      & $6.76\ 10^{1}$ \\ \cline{2-3}
                             & $B_2^+\longrightarrow B^{*+}\gamma$      & $2.70\ 10^{2}$ \\ \hline
$2^+\longrightarrow 0^-+1^-$ & $B_2^0\longrightarrow B^0\gamma$         & $4.08        $ \\ \cline{2-3}
                             & $B_2^+\longrightarrow B^+\gamma$         & $1.63\ 10^{1}$ \\ \hline
$2^+\longrightarrow 1^++1^-$ & $B_{s2}^0\longrightarrow B_{s1}^0\gamma$ & $3.38\ 10^{-4}$ \\ \hline
$2^+\longrightarrow 1^-+1^-$ & $B_{s2}^0\longrightarrow B_s^{*0}\gamma$ & $5.86\ 10^{1}$ \\ \hline
$2^+\longrightarrow 0^-+1^-$ & $B_{s2}^0\longrightarrow B_s^0\gamma$    & $4.62        $ \\ \hline
$1^+\longrightarrow 1^-+1^-$ & $B_1^0\longrightarrow B^{*0}\gamma$      & $2.73\ 10^{1}$ \\ \cline{2-3}
                             & $B_1^+\longrightarrow B^{*+}\gamma$      & $1.09\ 10^{2}$ \\ \hline
$1^+\longrightarrow 0^-+1^-$ & $B_1^0\longrightarrow B^0\gamma$         & $4.10\ 10^{1}$ \\ \cline{2-3}
                             & $B_1^+\longrightarrow B^+\gamma$         & $1.64\ 10^{2}$ \\ \hline
$1^+\longrightarrow 1^-+1^-$ & $B_{s1}^0\longrightarrow B_s^{*0}\gamma$ & $2.56\ 10^{1}$ \\ \hline
$1^+\longrightarrow 0^-+1^-$ & $B_{s1}^0\longrightarrow B_s^0\gamma$    & $3.46\ 10^{1}$ \\ \hline
$1^-\longrightarrow 0^-+1^-$ & $B^{*0}\longrightarrow B^0\gamma$        & $6.41\ 10^{-2}$ \\ \cline{2-3}
                             & $B^{*+}\longrightarrow B^+\gamma$        & $2.72\ 10^{-1}$ \\ \cline{2-3}
                             & $B_s^{*0}\longrightarrow B_s^0\gamma$    & $5.10\ 10^{-2}$ \\ \cline{2-3}
			     & $D^{*0}\longrightarrow D^0\gamma$        & $7.18         $ \\ \cline{2-3}
                             & $D^{*+}\longrightarrow D^+\gamma$        & $1.73         $ \\ \hline\hline
\end{tabular}
\caption{Photon decays}
\label{tab:ratetab}
\end{center}
\end{table}

 It is also possible to calculate the coupling 
$g_{B^*B\pi}$, the $B^*$ emits a virtual pion. This coupling has been 
calculated in the rest system of the heavy meson at zero recoil, 
the result is :
\begin{align}
g_{B^*B\pi^\pm}&=\sqrt{\fr{2}{9}}\fr{R}{f_\pi} 
|R^2\left( f_\beta^*(R)g_\alpha (R)+
g_\beta^*(R)f_\alpha (R)\right)|M_{B^*}\\
g_{B^*B\pi^0}&=\sqrt{\fr{1}{9}}\fr{R}{f_\pi} 
|R^2\left( f_\beta^*(R)g_\alpha (R)+
g_\beta^*(R)f_\alpha (R)\right)|M_{B^*}.\\
\intertext{Using the 
wavefunctions for the $B^*$ and $B$ meson give us}
g_{B^*B\pi^+}&=45.6 \\
g_{B^*B\pi^0}&=32.2. 
\end{align}
The coupling of photons to the mesonic states are done straight forward 
by using the interaction Lagrangian :
\begin{equation}
{\cal L}_{\text{int}}=e_qe\psibar\bs{\gamma\cdot A}\psi
\end{equation}
Some expressions for electromagnetic transitions are listed below :
\begin{align}
0^-\longrightarrow 2^+1^-&
\begin{cases}\Gamma=&\fr{24}{5}\alpha e_q^2\wk
\left|{\int dx}j_2(F+G)\right|^2 
\end{cases}\label{eq:2sg} \\
0^-\longrightarrow 1^+1^-&
\begin{cases}
\Gamma=&\fr{8}{3}\alpha e_q^2\wk
\left(\right.\left.\fr{1}{5}\left|{\int dx}j_2(F+G)
\right|^2\right. \\
&\left.+\left|{\int dx}(Fj_2-Gj_0)\right|^2\right. 
\left.+\fr{1}{3}\left|{\int dx}(Gj_2+Gj_0)\right|^2\right)
\end{cases} \\
0^-\longrightarrow 1^-1^-&
\begin{cases}\Gamma=&\fr{24}{5}\alpha e_q^2\wk
\left|{\int dx}j_1(F+G)\right|^2 
\end{cases} \\
0^-\longrightarrow 0^-1^-&
\begin{cases}\Gamma=&0
\end{cases}\\
2^+\longrightarrow 1^+1^-&
\begin{cases}
\Gamma=&\fr{2}{375}\alpha e_q^2\wk\left(\fr{87}{4}\left|{\int dx}j_1(F+G)
\right|^2\right. \\&\left.+52\left|{\int dx}(Gj_3+Gj_1)\right|^2
+2\left|{\int dx}(Gj_3-Fj_1)\right|^2\right. \\
&\left.+27\left|{\int dx}(Fj_3-Gj_1)\right|^2
+27\left|{\int dx}(Fj_3+Fj_1)\right|^2\right. \\
&\left.+\fr{3501}{7}\left|{\int dx}j_3(F+G)\right|^2\right)
\end{cases} \\
2^+\longrightarrow 1^-1^-&
\begin{cases}
\Gamma=&\fr{4}{3}\alpha e_q^2\wk
\left(\fr{11}{10}\left|{\int dx}j_2(F+G)\right|^2
+\left|{\int dx}(Gj_2-Fj_0)\right|^2\right. \\
&\left.+\fr{1}{3}\left|{\int dx}(Fj_2+Fj_0)\right|^2\right)
\end{cases}\label{eq:p3/22}\\
2^+\longrightarrow 0^-1^-&
\begin{cases}\Gamma=&\fr{6}{5}\alpha e_q^2\wk
\left|{\int dx}j_2(F+G)\right|^2 
\end{cases}\label{eq:p3/23} \\
1^+\longrightarrow 1^-1^-&
\begin{cases}
\Gamma=&\fr{2}{9}\alpha e_q^2\wk\left(13\left|{\int dx}j_2(F+G)\right|^2
+\fr{2}{3}\left|{\int dx}(Fj_2+Fj_0)\right|^2\right. \\
&\left.+2\left|{\int dx}(Gj_2-Fj_0)\right|^2\right)
\end{cases}\label{eq:p3/21}\\
1^+\longrightarrow 0^-1^-&
\begin{cases}
\Gamma=&\fr{8}{9}\alpha e_q^2\wk\left(-\fr{1}{4}\left|{\int dx}j_2(F+G)
\right|^2+\fr{1}{3}\left|{\int dx}(Fj_2+Fj_0)\right|^2\right. \\
&\left.+\left|{\int dx}(Gj_2-Fj_0)\right|^2\right)
\end{cases}\label{eq:1sg}\\
1^-\longrightarrow 0^-1^-&
\begin{cases}\Gamma=&\fr{4}{3}\alpha e_q^2\wk
\left|{\int dx}j_1(F+G)\right|^2 
\end{cases}\label{eq:p3/20}
\end{align}
where we have defined :
\begin{align}
F&\ee x^2{f_\beta (x)}^* g_\alpha (x)\label{eq:transf} \\
G&\ee x^2{g_\beta (x)}^* f_\alpha (x)\label{eq:transg} \\
j_l&\ee j_l(\wk x)
\end{align}

The numerical values for the above expressions
are shown in table \ref{tab:ratetab}.
The transitions in table \ref{tab:ratetab} are very suppressed in 
comparison with those listed in table \ref{tab:ratetab2}.
This of course is expected,
the smaller phase space for pion decays is compensated by the much 
stronger pion coupling relative to the electromagnetic coupling.

\subsection{Comparison with theoretical and experimental results}

We have calculated a lot of partial widths for different particles,
listed in table  \ref{tab:ratetab2} and \ref{tab:ratetab}. To day very little
is known on the experimental front, but there are a lot 
of theoretical predictions. So we will compare our results with the 
known experimental and some of the theoretical results. As we shall see 
there are no conflict between our predictions and the experimental
information.

In table \ref{tab:teor} we have listed some theoretical and experimental
results, on the experimental limits we have assumed that the
\begin{table}[htb]
$$
\begin{array}{|l|c|c|c|c|c|}\hline\hline
 &\multicolumn{4}{|c|}{\text{Theoretical}}&\text{Experimental}\\ \cline{2-6}
 &\cite{quigg2}(MeV)&\cite{donell}(keV)&\cite{zhu}(keV)&\text{our work}&\cite{rev} (keV)\\ \hline
\Gamma(B_2\longr B^*\pi)       &11 &-   &-             &13.8MeV &- \\ \hline
\Gamma(B_2\longr B\pi)         &10 &-   &-             &14.7MeV &- \\ \hline
\Gamma(B_1\longr B^*\pi)       &14 &-   &-             &20.0MeV &- \\ \hline
\Gamma(B^{*+}\longr B\gamma)   &-  &-   &0.38\pm 0.06  &0.272keV &- \\ \hline
\Gamma(B^{*0}\longr B^0\gamma) &-  &-   &0.13\pm 0.03  &6.41\ 10^{-2}keV  &- \\ \hline
\Gamma(B_s^*\longr B_s\gamma)    &-  &-   &0.22\pm 0.04  &5.10\ 10^{-2}keV  &-\\ \hline
\Gamma(D^{*+}\longr D^0\pi^+)  &-  &69.1&-             &60.1keV & <91 \\ \hline
\Gamma(D^{*+}\longr D^+\pi^0)  &-  &32.1&-             &27.2keV & <44 \\ \hline
\Gamma(D^{*0}\longr D^0\pi^0)  &-  &46.0&-             &38.8keV & <85 \\ \hline
\Gamma(D^{*+}\longr D^+\gamma) &-  &0.919&0.23\pm 0.1  &1.72keV & <4.1\\ \hline
\Gamma(D^{*0}\longr D^0\gamma) &-  &23.5&12.9\pm 2     &7.18keV & <54  \\ \hline\hline
\end{array}
$$
\caption{Comparison with theoretical and experimental results}\label{tab:teor}
\end{table}
the width of $D^{*0}$ have the 
same upper limit as the width of $D^{*\pm}$. It may not be clear from table 
\ref{tab:teor}, but the theoretical predictions vary a lot.
In \cite{belyaev} there is a summary 
of theoretical estimates. For the particular decay 
$D^{*+}\longr D^0\pi^+$ that determines the coupling constant 
$g_{D^*D\pi}$, the predicted 
rates vary from $10keV$(QCD sum rules) to more than $100keV$(quark model +
chiral HQET). We obtained $\Gamma(D^{*+}\longr D^0\pi^+)=60.1keV$, close to 
the value ($61-78keV$) coming from P.Cho and H.Georgi \cite{cho} who makes calculations 
with chiral HQET. The value of the coupling $g_{B^*B\pi^+}$
vary from $g_{B^*B\pi^+}=15\pm 4$(QCD sum rules) to $g_{B^*B\pi^+}=64$(quark
model+chiral HQET), we found $g_{B^*B\pi^+}=45.6$.

We have calculated most but not all decay modes for the excited states. 
The $\pi\pi$ modes are missing. As we believe that these modes are less 
important than the emission of single pions in the decays we still can 
give approximate values of the decay widths, these are : 
\begin{alignat}{3}
&2S_{1/2}\ :&\quad&\Gamma(B^\p)\simeq 195MeV & &\\
&2P_{3/2}\ :&\quad&\Gamma(B_2)\simeq 29MeV &\qquad &\Gamma(B_1)\simeq 20MeV\\
&2P_{1/2}\ :&\quad&\Gamma(B_1)\simeq 139MeV &\qquad &\Gamma(B_0)\simeq 141MeV \\
&1S_{1/2}\ :&\quad&\Gamma(D^{*+})\simeq 89keV &\qquad 
&\Gamma(D^{*0})\simeq 46keV \label{eq:widthd}
\end{alignat}

The $P_{1/2}$ states are naturally much wider than the $P_{3/2}$ states 
because they decay only trough an $S$-wave,
whereas the $P_{3/2}$ states decay through a $D$-wave. The full width of 
the $P$ states indicated by a preliminary experiment \cite{delphi}  
are $\Gamma(B(P_{3/2}))\simeq 20MeV$
and $\Gamma(B(P_{1/2}))\simeq 150MeV$. This is, as we see, in good agreement 
with our results. Since the $P_{1/2}$ states are so broad, they are very 
hard to reconstruct from the experimental results, and so far there have not
been any really precise measurement of their masses.

The CLEO report \cite{cleo2} contains best
measurement of the $D^{*+}$ 
branching fractions, a large improvement of what can be found in the Particle 
Data Book \cite{rev}. Since we have calculated the width of the $D^{*+}$ meson, 
equation (\ref{eq:widthd}), it 
is easy to calculate the branching ratios. The results are shown in table 
\ref{tab:cleo} together with the CLEO results.
\begin{table}[htb]
$$
\begin{array}{|l|c|c|}\hline\hline
&\text{CLEO}&\text{our calculations}\\ \hline
B_r(D^{*+}\longr D^+\gamma)&(1.68\pm0.51)\% &1.94\% \\ \hline
B_r(D^{*+}\longr D^+\pi^0)&(30.73\pm0.63)\% &30.55\% \\ \hline
B_r(D^{*+}\longr D^+\pi^+)&(67.59\pm0.70)\% &67.51\% \\ \hline\hline
\end{array}
$$
\caption{Comparison with CLEO data and our predictions}\label{tab:cleo}
\end{table}
Our results are clearly in good agreement with the experimental data
We recognize however that branching ratios are one thing, 
particular decay widths another. We get the correct ratio between 
pionic and electromagnetic decays.
As the coupling of the electromagnetic field to the quarks is simple, we 
naturally have some confidence in the calculated electromagnetic transitions 
rates. Therefore we believe that our calculated pionic rates cannot be too
far off from what will be measured in the future.

\bibliographystyle{unsrt}

\begin{thebibliography}{99}
\bibitem{isgur2} Nathan Isgur and Mark B.Wise, Phys. Lett. {\bf B 232}, 113, 1989.
\bibitem{isgur3} Nathan Isgur and Mark B.Wise, Phys. Lett. {\bf B 237}, 527, 1990.
\bibitem{georgi1} H.Georgi, Phys.Lett. {\bf B 240}, 447, 1990.
\bibitem{eichten1} E.Eichten and B.Hill, Phys. Lett. {\bf B 234}, 511, 1990.
\bibitem{eichten2} E.Eichten and B.Hill, Phys. Lett. {\bf B 243}, 259, 1990.
\bibitem{neubert1} M.Neubert, Phys.Rep. {\bf 245}, 259, 1994.
\bibitem{wilcox} Walter Wilcox and O. V. Maxwell and Kimball A. Milton, Phys.Rev.
{\bf D 31}, 1081, 1985.
\bibitem{hog2} H.Hogaasen and M.Sadzikowski, Z. Phys. {\bf C 64}, 427, 1994.
\bibitem{sad2} M. Sadzikowski and K. Zalewski, Z. Phys. {\bf C 59}, 677, 1993.
\bibitem{bog} P. N. Bogolioubov, Ann. Inst. Henri Poincar\'{e} {\bf VIII}, 163, 1968.
\bibitem{chodos:1} A. Chodos and R. L. Jaffe and K. Johnson C. B. Thorn and V. F. Weisskopf, Phys.Rev. {\bf D 9}, 3471, 1974
\bibitem{hog} H. Hogaasen and J. M. Richard and P. Sorba, Phys.Lett.
{\bf B 119}, 272
\bibitem{milton} Kimball A. Milton and Lester L. DeRaad Jr. and 
Julian Schwingerm, Ann. of Phys. {\bf 115}, 388, 1978.
\bibitem{detar} Dale Izatt and Carleton Detar and Mark Stephenson
Nucl.Phys. {\bf B 199}, 269, 1982.
\bibitem{rev} R. M. Barnett et al.,  Phys.Rev. {\bf D 54}, 1.
\bibitem{opal} OPAL Collaboration, R. Akers et al., Z.Phys. {\bf C 66}, 19,
1995.
\bibitem{weiser} C. Weiser, Talk given at the {XXVII} International Conference
on High Energy Physics, Warsaw, 25-31 July 1996.
\bibitem{thomas} A. W. Thomas Advances in Nuclear Physics, {\bf 13}, 1, 1983.
\bibitem{quigg2} Estia J. Eichten and Christopher T. Hill and Chris Quigg, 
Phys. Rev. Lett. {\bf 71}, 4116, 1993.
\bibitem{donell} Patric J.O'{D}onnel and Q.P.Xu, Phys.Lett. {\bf B 336}, 113, 1994.
\bibitem{zhu} Shi-lin Zhu and W-{Y}.P.Hwang and Ze-sen Yang, {\bf hep-ph/9610412}, 1997.
\bibitem{belyaev} V.M.Belyaev and V.M.Braun and A.Khodjamirian and R.R\"{u}ckl,
Phys.Rev {\bf D 51}, 6177, 1995.
\bibitem{cho} Peter Cho and Howard Georgi, Phys.Lett. {\bf B 296}, 408, 1992.
\bibitem{delphi} {DELPHI} {C}ollaboration, P. Abreu et al., 
Phys.Lett. {\bf B 345}, 598, 1995.
\bibitem{cleo2} J.Bartelt et al., Phys.Rev.Lett, 3919, 1998.
\end{thebibliography}

\end{document}